\documentclass[aps,prl,twocolumn,superscriptaddress]{revtex4}
\usepackage{amssymb}
\usepackage{amsmath}
\usepackage{epsfig}
\usepackage{color}
\usepackage{graphics, graphicx}
\usepackage{bbold}
\usepackage{psfrag}
\usepackage{subfigure}
\usepackage{verbatim}
\usepackage{color}
\usepackage[colorlinks,citecolor=blue]{hyperref}

\begin{document}

\title{Rashba and Weyl spin-orbit coupling in an optical lattice clock}
\author{Xiaofan Zhou}
\affiliation{Department of Physics, The University of Texas at Dallas, Richardson, Texas
75080, USA}
\affiliation{State Key Laboratory of Quantum Optics and Quantum Optics Devices, Institute
of Laser spectroscopy, Shanxi University, Taiyuan 030006, China}
\author{Xi-Wang Luo}
\affiliation{Department of Physics, The University of Texas at Dallas, Richardson, Texas
75080, USA}
\author{Gang Chen}
\email{chengang971@163.com}
\affiliation{State Key Laboratory of Quantum Optics and Quantum Optics Devices, Institute
of Laser spectroscopy, Shanxi University, Taiyuan 030006, China}
\affiliation{Collaborative Innovation Center of Extreme Optics, Shanxi University,
Taiyuan, Shanxi 030006, China}
\author{Suotang Jia}
\affiliation{State Key Laboratory of Quantum Optics and Quantum Optics Devices, Institute
of Laser spectroscopy, Shanxi University, Taiyuan 030006, China}
\affiliation{Collaborative Innovation Center of Extreme Optics, Shanxi University,
Taiyuan, Shanxi 030006, China}
\author{Chuanwei Zhang}
\email{chuanwei.zhang@utdallas.edu}
\affiliation{Department of Physics, The University of Texas at Dallas, Richardson, Texas
75080, USA}

\begin{abstract}
Recent experimental realization of one-dimensional (1D) spin-orbit coupling
(SOC) for ultracold alkaline-earth(-like) atoms in optical lattice clocks
opens a new avenue for exploring exotic quantum matter because of the
strongly suppressed heating of atoms from lasers comparing with alkaline
atoms. Here we propose a scheme to realize two-dimensional (2D) Rashba and
three-dimensional (3D) Weyl types of SOC in a 3D optical lattice clock and
explore their topological phases. With 3D Weyl SOC, the system can support
topological phases with various numbers as well as types (I or II) of Weyl
points. The spin textures of such topological bands for 2D Rashba and 3D
Weyl SOC can be detected using suitably designed spectroscopic sequences.
Our proposal may pave the way for the experimental realization of robust
topological quantum matters and their exotic quasiparticle excitations in
ultracold atomic gases.
\end{abstract}

\maketitle

\emph{Introduction}.---Spin-orbit coupling (SOC) plays a key role for many
condensed matter phenomena, such as anomalous and spin Hall effects \cite%
{Xiao}, topological insulators and superconductors~\cite{Hasan,Qi,Moore},
etc. The recent experimental realization of 1D~\cite%
{spielman,FuPRA,Shuai-PRL,Washington-PRA,Purdue,Jing,MIT,Spielman-Fermi} and
2D~\cite{2DSOCpan,2DSOChuang,2DSOCmeng} SOC in ultracold alkaline atoms
provides a highly controllable and disorder-free platform for exploring
nontrivial topological physics induced by SOC, such as Majorana fermions
with non-Abelian exchange statistics~\cite%
{Majorana2008,Majorana2009,Zhang2008,Majorana2012} and Weyl fermions \cite%
{Weyl1,Weyl2,Weyl3,Weyl4,Weyl5} carrying topological monopole charges~\cite%
{Monopole}. However, the experimental observation of these topological
phenomena is greatly hindered by the heating of atoms, particularly
fermions, originating from the Raman process where lasers couple hyperfine
ground states with high-lying excited states. The single-photon detuning for
Raman lasers is limited by the fine-structure splitting~\cite%
{spielman,FuPRA,Shuai-PRL,Washington-PRA,Purdue,Jing,MIT,Spielman-Fermi,2DSOCpan,2DSOChuang,2DSOCmeng}
for the generation of SOC, which is usually small for alkaline atoms,
yielding large spontaneous emission of photons that heat the atomic gas.

The heating issue may be overcome by choosing atomic species with large
fine-structure splitting such as Dy and Er~\cite{lanth1,lanth2} or using
alkaline-earth(-like) atoms~\cite{clocktheory} with long-lived excited
states [e.g., the lifetime for $^{87}$Sr ($^{173}$Yb) is $\sim $160 ($\sim $%
20 ) seconds]. For alkaline-earth(-like) atoms, 1D SOC has been
theoretically proposed \cite{clocktheory} and experimentally realized \cite%
{clock87Sr,clock173Yb,clocksoc} recently through directly coupling the
ground $^{1}S_{0}$ (referred as $|g\rangle $) and excited metastable $%
^{3}P_{0}$ (referred as $|e\rangle $) clock-states in 1D optical lattice
clocks, which does not involve any Raman process. With the recent
experimental success in realizing 2D and 3D optical lattice clocks for both
Bose and Fermi atoms \cite{3Dclockbose,3Dclockfermi}, a natural question is
whether 2D and 3D SOC can also be realized without any Raman process.

\begin{figure}[t]
\centering
\includegraphics[width = 3.4in]{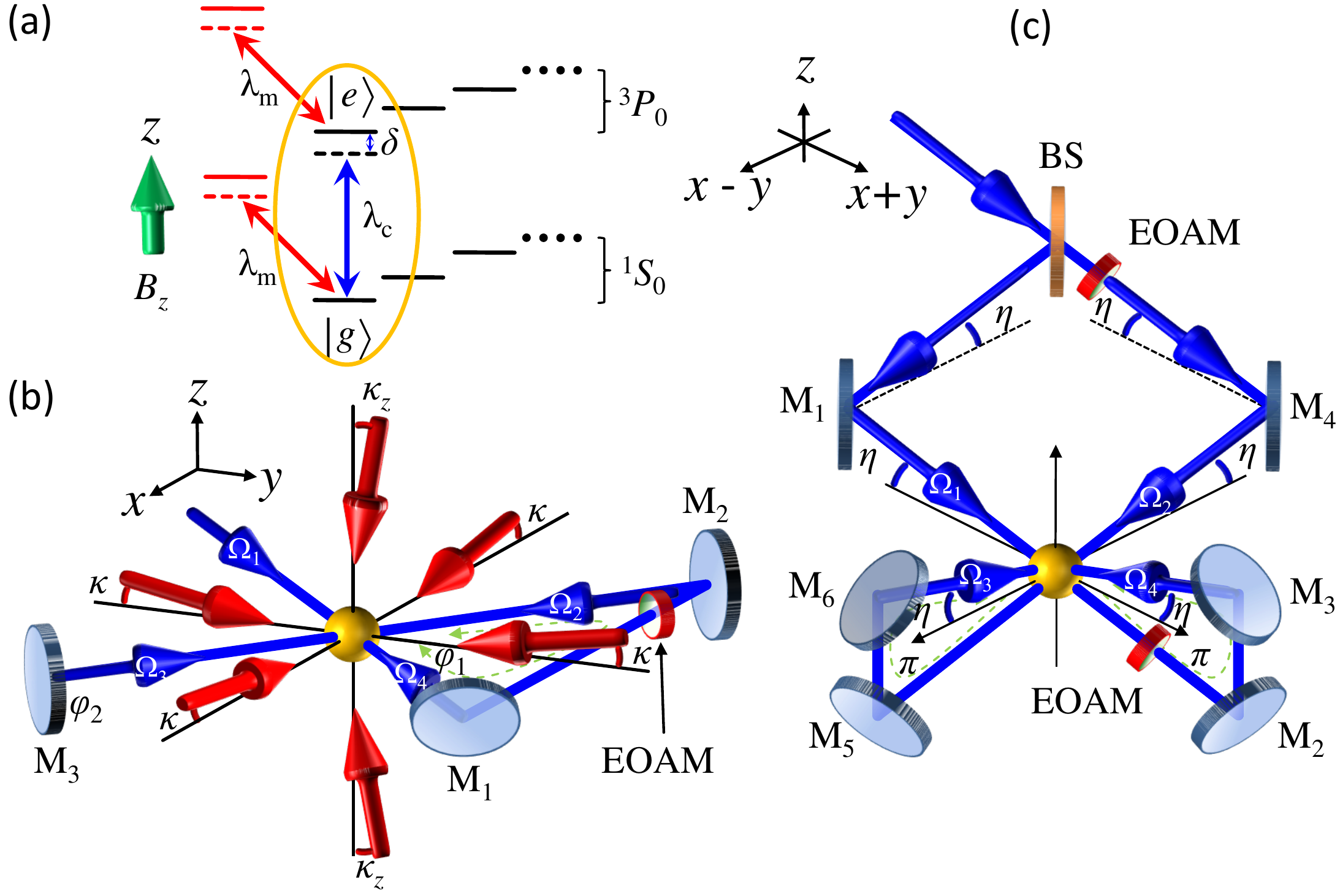} \hskip 0.0cm
\caption{Schematics of the proposed experimental setup for generating 2D
Rashba and 3D Weyl SOC. (a) Optical transitions. With a magnetic field
applied along the \textit{z}-direction, two nuclear-spin-polarized states
can be isolated and chosen as two spin states $|g\rangle $ and $|e\rangle $
for the SOC. Blue arrow represents the clock laser transition, with the
single-photon detuning $\protect\delta $. Red arrows represent magic
wavelength lasers for 3D optical lattices. (b) Laser setup for 2D Rashba
SOC. (c) Clock laser setup for 3D Weyl SOC. Optical lattice lasers are the
same as (b). Here EOAM represents electro-optic amplitude modulator which is
used to turn on/off relevant laser beams for the experimental detection. }
\label{fig:experimental}
\end{figure}

Here, we address this important question by proposing a scheme for realizing
both 2D Rashba and 3D Weyl types of SOC for alkaline-earth(-like) atoms in a
3D optical lattice clock~\cite{clockreview} using a simple experimental
setup without involving Raman process. The experimental realization of the
proposed scheme should pave the way for the eventual experimental generation
of stable topological superfluids without heating and the observation of
topological Majorana \cite{Zhang2008,Majorana2012} and Weyl fermions \cite%
{Weyl1,Weyl2,Weyl3} in ultracold atomic gases. Our main results are:

\textit{i}) Beside the 3D optical lattice potential generated with magic
wavelength lasers, a clock laser that couples two states $|g\rangle $ and $%
|e\rangle $ is coherently splitted into four beams propagating along
different directions (see Fig. \ref{fig:experimental}) and their
interference generates the SOC. 3D Weyl (2D Rashba) types of SOC are
realized when the wavevectors of the clock laser do (not) possess \textit{z}%
-components.

\textit{ii}) In the presence of 3D Weyl types of SOC, there exists a rich
phase diagram containing topological phases with various number of Weyl
points as well as a fully gapped 3D Chern insulator phase. The Weyl points
can be type-I or type-II \cite{Weyl2,typeII}. Different Weyl points with
opposite topological charges are connected by gapless Fermi arcs on the
surface.

\textit{iii}) Three spectroscopic sequences are proposed to accurately
measure the spin textures of the topological bands for 2D Rashba and 3D Weyl
SOC using a combination of Rabi spectroscopy and time-of-flight images.

\emph{Experimental scheme and theoretical modeling}.---Our proposed
experimental scheme for generating SOC for alkaline-earth(-like) atomic
gases $^{87}$Sr ($^{173}$Yb) \cite{clocktheory,clock87Sr,clock173Yb,clocksoc}%
~is illustrated in Fig.~\ref{fig:experimental}. A large magnetic field is
applied along the $z$-direction so that only two nuclear-spin-polarized
states ($|g\rangle $ and $|e\rangle $ that form an effective spin-1/2) in
ground $^{1}S_{0}$ and metastable excited $^{3}P_{0}$ manifolds are
populated and coupled by a clock laser [see Fig.~\ref{fig:experimental}(a)].
The state-independent 3D optical lattice potential $V_{\mathrm{lat}}\left( 
\mathbf{r}\right) =-V_{0}\left[ \cos ^{2}\left( k_{\mathrm{L}}x\right) +\cos
^{2}\left( k_{\mathrm{L}}y\right) \right] -V_{z}\cos ^{2}\left( k_{\mathrm{L}%
}^{z}z\right) $ (the same for $|g\rangle $ and $|e\rangle $) are implemented
using six plane-wave lasers [see red arrows in Fig.~\ref{fig:experimental}%
(b)] with the magic wavelength $\lambda _{\mathrm{m}}$ ($813$ nm for $^{87}$%
Sr and $759$ nm for $^{173}$Yb) \cite{clockreview}. Here $k_{\mathrm{L}%
}=2\pi \cos \left( \kappa \right) /\lambda _{\mathrm{m}}$ and $k_{\mathrm{L}%
}^{z}=2\pi \cos \left( \kappa _{z}\right) /\lambda _{\mathrm{m}}$ are the
wavevectors in the $x$-$y$ plane and $z$-direction, with $\kappa $ and $%
\kappa _{z}$ corresponding laser incident angles.

The two states $|g\rangle $ and $|e\rangle $ are coupled by a clock laser
with wavelength $\lambda _{\mathrm{c}}$ ($698$ nm for $^{87}$Sr and $578$ nm
for $^{173}$Yb). In the rotating frame, the single-particle Hamiltonian can
be written as 
\begin{equation}
H=\left[ \frac{\mathbf{p}^{2}}{2m}+\!V_{\mathrm{lat}}\left( \mathbf{r}%
\right) \right] \!\!I+m_{z}\sigma _{z}+\hbar \left( M\!\sigma _{+}\!+\!%
\mathrm{H.c.}\right) ,  \label{Ham1}
\end{equation}%
where $\mathbf{p}$ is the momentum operator, $m$ is the mass of atoms, $%
m_{z}=\hbar \delta $ is the effective Zeeman field determined by the clock
laser detuning, and $\sigma _{j}$ ($I$) is the Pauli (identity) matrix on
the $\left\{ \left\vert g\right\rangle ,\left\vert e\right\rangle \right\} $
basis. For the generation of 2D and 3D SOC, the spatial distribution of the
Rabi coupling of the clock laser is designed to be $M=M_{0}e^{-ik_{\mathrm{R}%
}^{z}z}\left[ \sin (k_{\mathrm{c}}x)\cos (k_{\mathrm{c}}y)+i\cos (k_{\mathrm{%
c}}x)\sin (k_{\mathrm{c}}y)\right] $, which can be realized through suitable
interference of the clock laser beams. Figure~\ref{fig:experimental}(b)
shows the experimental setup for realizing 2D Rashba SOC with $k_{\mathrm{R}%
}^{z}=0$, where the clock laser (linearly polarized along the $z$-direction)
is reflected by three mirrors ($\text{M}_{1}$, $\text{M}_{2}$ and $\text{M}%
_{3}$) and propagates along the ($x\pm y$)-directions in the intersecting
area with corresponding Rabi frequencies $\Omega _{1}=\Omega _{0}e^{ik_{%
\mathrm{c}}\left( x+y\right) }$, $\Omega _{2}=\Omega _{0}e^{ik_{\mathrm{c}%
}\left( x-y\right) +i\varphi _{1}}$, $\Omega _{3}=\Omega _{0}e^{-ik_{\mathrm{%
c}}\left( x-y\right) +i\left( \varphi _{1}+\varphi _{2}\right) }$, and $%
\Omega _{4}=\Omega _{0}e^{-ik_{\mathrm{c}}\left( x+y\right) +i\left(
2\varphi _{1}+\varphi _{2}\right) }$. Without loss of generality, we set $%
\Omega _{0}$ to be real because its overall phase originating from the
initial phase of the incident laser can be gauged out without affecting the
SOC. $k_{\mathrm{c}}=2\pi /\left( \sqrt{2}\lambda _{\mathrm{c}}\right) $, $%
\varphi _{1}$ ($\varphi _{2}$) is the phase acquired by the beam when it
propagates along the optical path from the atom cloud over mirrors $\text{M}%
_{1,2}$ ($\text{M}_{3}$) then back to the atom cloud. The total Rabi
coupling strength $M=\sum_{\Lambda =1}^{4}\Omega _{\Lambda }$ has above
designed form by choosing $\varphi _{1}\!\!=\!\!0$ (mod 2$\pi $), $\varphi
_{2}\!\!=\!\!\pi $ (mod 2$\pi $), and $M_{0}=2\sqrt{2}\Omega _{0}$. We
choose $k_{\mathrm{c}}=k_{\mathrm{L}}$ for the generation of desired 2D
Rashba SOC, yielding $\cos \left( \kappa \right) \!=\!\lambda _{\mathrm{m}%
}/\!\left( \sqrt{2}\lambda _{\mathrm{c}}\right) \approx \!0.8$ for $^{87}$Sr
($0.9$ for $^{173}$Yb).

The generation of 3D SOC requires the phase factor $e^{-ik_{\mathrm{R}%
}^{z}z} $ in the Rabi coupling $M$, which can be realized by tilting the
four clock laser beams $\Omega _{\Lambda }$ by an angle $\eta $ with respect
to the $x$-$y$ plane with $k_{\mathrm{R}}^{z}=2\pi \sin (\eta )/\lambda _{%
\mathrm{c}}$. In the $x$-$y$ plane, $k_{\mathrm{L}}=k_{\mathrm{c}}=2\pi \cos
(\eta )/\left( \sqrt{2}\lambda _{\mathrm{c}}\right) $ yields $\cos \left(
\kappa \right) \!/\!\cos \left( \eta \right) \!=\!\lambda _{\mathrm{m}%
}/\!\left( \sqrt{2}\lambda _{\mathrm{c}}\right) $. Such 3D $\Omega _{\Lambda
}$ can be realized with a similar optical setup with mirrors and a beam
splitter [see Fig.~\ref{fig:experimental}(c)]. Note that the electric field
of the clock laser has a component in the $x$-$y$ plane, which, however,
does not induce the transition to other nuclear-spin states due to the large
Zeeman splitting [see Fig.~\ref{fig:experimental}(a)]. Such Rabi coupling
between two pseudospin states does not involve the Raman process that
requires a small detuning to unstable high excited states and induces large
spontaneous emission of photons. Furthermore, the Raman transition between
different Zeeman states requires careful design of laser polarizations of
Raman beams, which is not necessary for the Rabi transition here, yielding a
simpler optical setup.

With typical optical lattice potential depths, atoms are confined in the
lowest band of the lattice and tight-binding approximation can be applied.
It is straightforward to derive corresponding tight-binding lattice model
for the Hamiltonian Eq.~(\ref{Ham1}), which yields an effective Hamiltonian 
\cite{supp} 
\begin{equation}
\mathcal{H}_{\mathrm{E}}=\sum\nolimits_{\mathbf{k},ss^{\prime }}\hat{c}_{%
\mathbf{k},s}^{\dag }\mathbf{H}_{\mathbf{k}}\hat{c}_{\mathbf{k},s^{\prime }},
\label{Ham2}
\end{equation}%
where $s=\left( g,e\right) $, and $\hat{c}_{\mathbf{k},s}^{\dag }$ ($\hat{c}%
_{\mathbf{k},s}$) is the creation (annihilation) operator for state $s$ at
momentum $\mathbf{k}=\left( k_{x},k_{y},k_{z}\right) $. $\mathbf{H}_{\mathbf{%
k}}=h_{0\mathbf{k}}I+\mathbf{h}_{\mathbf{k}}\cdot \mathbf{\sigma }$, where $%
h_{0\mathbf{k}}=-2t_{z}\cos \left( k_{z}a_{z}\right) \cos \left( \phi
/2\right) $, $h_{x\mathbf{k}}=-2t_{\mathrm{so}}\sin (k_{y}a)$, $h_{y\mathbf{k%
}}=-2t_{\mathrm{so}}\sin (k_{x}a)$, and $h_{z\mathbf{k}}=m_{z}-2t\cos
(k_{x}a)-2t\cos (k_{y}a)+2t_{z}\sin \left( k_{z}a_{z}\right) \sin \left(
\phi /2\right) $. $t_{z}$ is the hopping parameter along the $z$-direction,
and $\phi =\pi k_{\mathrm{R}}^{z}/k_{\mathrm{L}}^{z}$. $a_{z}$ and $a$ are
optical lattice constants in the $z$-direction and $x$-$y$ plane,
respectively. $t$ and $t_{\mathrm{so}}$ are spin-preserved and spin-flipped
hopping parameters in the $x$-$y$ plane, and the latter is induced by the
Rabi coupling $M$ of the clock laser. The energy spectrum of the Hamiltonian
can be easily obtained through diagonalizing the $\mathbf{H}_{\mathbf{k}}$,
yielding $E_{\mathbf{k}}=h_{0\mathbf{k}}\pm |\mathbf{h}_{\mathbf{k}}|$.

When $k_{\mathrm{R}}^{z}=0$ (\textit{i.e.}, clock laser is in the $x$-$y$
plane), $\phi =0$ and $h_{z\mathbf{k}}=m_{z}-2t\cos (k_{x}a)-2t\cos (k_{y}a)$%
, therefore there is no coupling between momentum $k_{z}$ and spin, leaving
2D Rashba type of SOC in the $x$-$y$ plane in the Hamiltonian Eq.~(\ref{Ham2}%
). In this case, the single particle physics is described by a topological
phase transition between a trivial insulator for $|m_{z}|\!>\!4t$ and a
topological insulator for $|m_{z}|\!<\!4t$, with the phase boundary $%
|m_{z}|\!=\!4t$~determined by $|\mathbf{h}_{\mathbf{k}}|=0$ \cite%
{Xiongjun2014,2DSOCpan}. Such 2D Rashba SOC not only mixes states $%
\left\vert g\right\rangle $ and $\left\vert e\right\rangle $, but also lifts
the band degeneracy with a single Fermi surface. In the presence of \textit{s%
}-wave pairing interaction between two states, the fermionic superfluid
pairing supports Majorana fermions inside vortex cores \cite%
{Zhang2008,Majorana2012}, which possess non-Abelian exchange statistics and
are a building block for fault-tolerant topological quantum computation \cite%
{Majorana2008,Majorana2009}.

\begin{figure}[t]
\centering
\includegraphics[width = 3.4in]{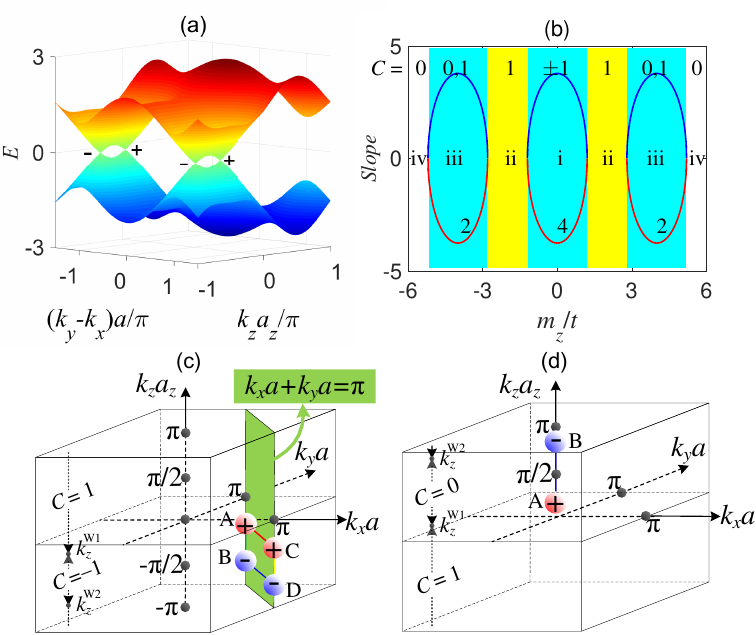} \hskip 0.0cm
\caption{Topological phases in 3D with $\protect\phi =\protect\pi $. In all
panels $t_{\mathrm{so}}/t=0.6$, $t_{z}/t=0.6$. (a) Energy spectrum $E_{%
\mathbf{k}}$ with four Weyl points in the $k_{x}a+k_{y}a=\protect\pi $
plane, with their topological charges $\pm 1$. $m_{z}/t=1$. (b) The phase
diagram with different $m_{z}$. $(2,4,2)$ indicate the number of Weyl
points. Blue and red lines represent two slopes $\pm 2t_{z}\cos \!\left( 
\protect\gamma \right) $ of the energy dispersions along the $k_{z}$%
-direction near a Weyl point. The Chern number $C$ is defined in the $k_{x}$-%
$k_{y}$ plane for fixed $k_{z}a_{z}$. (c) and (d) Positions and charges of
Weyl points for cases \textit{i}) and \textit{iii}) respectively. }
\label{fig:3D_SOC_z_lattice}
\end{figure}

\textit{Topological phases with 3D SOC}.---When $k_{\mathrm{R}}^{z}\neq 0$, $%
\phi =\pi k_{\mathrm{R}}^{z}/k_{\mathrm{L}}^{z}\neq 0$, and the Hamiltonian
Eq.~(\ref{Ham2}) contains the coupling $2t_{z}\sin \left( k_{z}a_{z}\right)
\sin \left( \phi /2\right) \sigma _{z}$ along the $k_{z}$-direction.
Hereafter we consider only $t>t_{z}$ and the topological phases for $t\leq
t_{z}$ can be discussed similarly and are presented in the supplementary
material \cite{supp}. We first consider $\phi =\pi $ (\textit{i.e.}, $k_{%
\mathrm{R}}^{z}=k_{\mathrm{L}}^{z}$), which leads to $h_{0\mathbf{k}}=0$.
From the energy dispersion $E_{\mathbf{k}}=\pm |\mathbf{h}_{\mathbf{k}}|$,
we see the spectrum is gapless only when $|\mathbf{h}_{\mathbf{k}}|=0$, with 
$h_{x\mathbf{k}}=h_{y\mathbf{k}}=0$ occurring at $k_{x}a,k_{y}a=0$ or $\pi $%
. At $\left( k_{x}a,k_{y}a\right) =\left( 0,\pi \right) $ or $\left( \pi
,0\right) $, $h_{z\mathbf{k}}=m_{z}+2t_{z}\sin \left( k_{z}a_{z}\right) $
becomes zero in the region $\left\vert m_{z}\right\vert \leq 2t_{z}$ at $%
k_{z}a_{z}=\gamma $ or $-\pi \!-\!\gamma $ with $\gamma =\arcsin \!\left[ 
\frac{-m_{z}}{2t_{z}}\right] $. While at $\left( k_{x}a,k_{y}a\right)
=\left( 0,0\right) $ or $\left( \pi ,\pi \right) $, $h_{z\mathbf{k}%
}=m_{z}\mp 4t+2t_{z}\sin \left( k_{z}a_{z}\right) $ becomes zero only in the
region $4t-2t_{z}\leq \pm m_{z}$ $\leq $ $4t+2t_{z}$ at $k_{z}a_{z}=\zeta $
or $\pi \!-\!\zeta $ with $\zeta =\arcsin \!\left[ \frac{\pm 4t-m_{z}}{2t_{z}%
}\right] $, where the top and bottom signs are for $\left( 0,0\right) $ and $%
\left( \pi ,\pi \right) $ respectively. Outside these two regions ($%
2t_{z}<\left\vert m_{z}\right\vert <4t-2t_{z}$ and $\left\vert
m_{z}\right\vert $ $>4t+2t_{z}$), the spectrum is fully gapped.

Each band gap closing point with $E_{\mathbf{k}}=0$ represents a Weyl point
[see Fig.~\ref{fig:3D_SOC_z_lattice}(a)], whose topological charge can be
determined by Chern number $\mathcal{C}=\frac{1}{2\pi }\oint_{S}B\left( 
\mathbf{k}\right) dS$ of the lowest energy band. Here $S$ is a surface
enclosing the Weyl point, and $B\left( \mathbf{k}\right) =i\left\langle
\nabla _{\mathbf{k}}\Psi \left( \mathbf{k}\right) \right\vert \times
\left\vert \nabla _{\mathbf{k}}\Psi \left( \mathbf{k}\right) \right\rangle $
is the Berry curvature~\cite{Xiao} with $\Psi \!\left( \mathbf{k}\right) $
the lower band wavefunction. Based on above gap closing conditions, there
exist four different phases for $t_{z}<t$ in different $m_{z}$ regions [see
Fig.~\ref{fig:3D_SOC_z_lattice}(b)]. \textit{i}) $\left\vert
m_{z}\right\vert \leq 2t_{z}$: four Weyl points at $\mathbf{k}^{W}=\left(
k_{x}a,k_{y}a,k_{z}a_{z}\right) =\left( 0,\pi ,\gamma \right) $, $\left(
0,\pi ,-\pi \!-\!\gamma \right) $, $\left( \pi ,0,\gamma \right) $ and $%
\left( \pi ,0,-\pi \!-\!\gamma \right) $ with different topological charges
[see Fig.~\ref{fig:3D_SOC_z_lattice}(c)]. For instance, around $\mathbf{k}%
^{W}=\left( 0,\pi ,\gamma \right) $, $\mathbf{h}_{\mathbf{k}}\approx 2t_{%
\mathrm{so}}a\left( \bar{k}_{x}\sigma _{x}-\bar{k}_{y}\sigma _{y}\right)
+2t_{z}a_{z}\cos \!\left( \gamma \right) \bar{k}_{z}\sigma _{z}$ with the
linear dispersion; \textit{ii}) $2t_{z}<\left\vert m_{z}\right\vert
<4t-2t_{z}$: a fully gapped topological phase with Chern number $C=1$ in the 
$k_{x}$-$k_{y}$ plane for any fixed $k_{z}$, corresponding to a stacking 2D
Chern insulator; \textit{iii}) $4t-2t_{z}\leq \pm m_{z}$ $\leq $ $4t+2t_{z}$%
: two Weyl points at $\mathbf{k}^{W}=\left( 0,0,\zeta \right) $ and $\left(
0,0,\pi \!-\!\zeta \right) $ for $m_{z}>0$ [see Fig.~\ref%
{fig:3D_SOC_z_lattice}(d)] or $\left( \pi ,\pi ,\zeta \right) $ and $\left(
\pi ,\pi ,\pi \!-\!\zeta \right) $ for $m_{z}<0$; \textit{iv}) $\left\vert
m_{z}\right\vert $ $>4t+2t_{z}$: trivial insulator phase.

\begin{figure}[t]
\includegraphics[width = 3.4in]{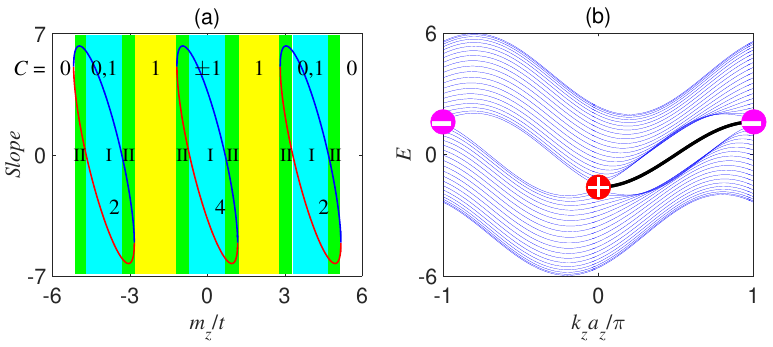} \hskip -0.0cm 
\centering
\caption{Type-I and type-II Weyl points for $\protect\phi \neq \protect\pi $%
. (a) Phase diagram. The notations and parameters are the same as those in
Fig.~\protect\ref{fig:3D_SOC_z_lattice}(b) except $\protect\phi =2\protect%
\pi /5$. I and II indicate the types of the Weyl points. Two slopes are $%
\protect\nu _{0}\pm \protect\nu _{z}$. (b) Energy spectrum $E_{\mathbf{k}}$
under open boundary condition along the $x$-direction with $\protect\phi =2%
\protect\pi /5$ and $m_{z}/t=0.0$. Black line represent the surface state
connecting two Weyl points. }
\label{fig:3D_SOC_z_lattice_E_k}
\end{figure}

For a general $\phi \neq \pi $, there is a nonzero $h_{0\mathbf{k}%
}=-2t_{z}\cos \left( k_{z}a_{z}\right) \cos \left( \phi /2\right) $ and $%
t_{z}$ in $h_{z\mathbf{k}}$ is also replaced by $t_{z}\!\sin \!\left( \phi
/2\right) $. $h_{0\mathbf{k}}$ does not change the eigenstates of the
Hamiltonian Eq.~(\ref{Ham2}), therefore the phase boundaries between above
four cases are only changed by the replacement $t_{z}\!\rightarrow
\!t_{z}\!\sin \!\left( \phi /2\right) $. However, nonzero $h_{0\mathbf{k}}$
rotates the slopes of the linear dispersions near the Weyl point such that
two slopes along the $k_{z}$-direction may have the same sign in certain
parameter region, which correspond to type-II Weyl points (the traditional
one with opposite signs of slopes is called type-I). For instance, around
the Weyl point $\mathbf{k}^{W}=\left( 0,\pi ,\gamma \right) $ for the case 
\textit{i}) $\left\vert m_{z}\right\vert \leq 2t_{z}\sin \!\left( \phi
/2\right) $, the Hamiltonian can be expanded as $\mathbf{h}_{\mathbf{k}}=\nu
_{0}\bar{k}_{z}+2t_{\mathrm{so}}a\left( \bar{k}_{x}\sigma _{x}-\bar{k}%
_{y}\sigma _{y}\right) +\nu _{z}\bar{k}_{z}\sigma _{z}$ with $\nu
_{0}=2t_{z}a_{z}\sin \!\left( \gamma \right) \cos \!\left( \phi /2\right) $, 
$\nu _{z}=2t_{z}a_{z}\cos \!\left( \gamma \right) \sin \!\left( \phi
/2\right) $, and $\gamma =\arcsin \!\left[ \frac{-m_{z}}{2t_{z}\sin \!\left(
\phi /2\right) }\right] $. The Lifshitz transition \cite{Lifshitz} between
type-I and type-II occurs at $\left\vert \nu _{0}\right\vert =\left\vert \nu
_{z}\right\vert $ [i.e., $\left\vert m_{z}^{c}\right\vert =2t_{z}\sin
^{2}\!\left( \phi /2\right) $]. The phase diagram and corresponding types of
Weyl points are showed in Fig.~\ref{fig:3D_SOC_z_lattice_E_k}(a). Finally,
because Weyl points do not stay at $E_{\mathbf{k}}=0$ plane due to nonzero $%
h_{0\mathbf{k}}$, the surface states are now embedded in the bulk spectrum
[see Fig.~\ref{fig:3D_SOC_z_lattice_E_k}(b)], instead of the straight Fermi
arc at $E_{\mathbf{k}}=0$ connecting two Weyl points with opposite charges
for $\phi =\pi $.

\begin{figure}[t]
\centering
\includegraphics[width = 3.4in]{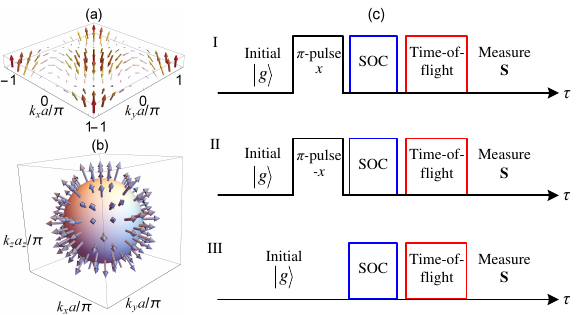} \hskip 0.0cm
\caption{Detection of spin textures. (a) Spin textures in the $k_{x}$-$k_{y}$
plane with 2D Rashba SOC. $t_{\mathrm{so}}/t=0.6$, $m_{z}/t=0$. (b) Spin
textures around a Weyl point with +1 topological charge (the arrows are
rotated $90^{\circ }$ along the $z$-axis for better illustration). $t_{%
\mathrm{so}}/t=0.6$, $t_{z}/t=0.6$, $m_{z}/t=0$, and $\protect\phi =\protect%
\pi $. (c) Three spectroscopic sequences I, II and III for the detection of
spin parameters $\protect\theta _{\mathbf{k}}$ and $\protect\varphi _{%
\mathbf{k}}$.}
\label{fig:detect}
\end{figure}

\emph{Experimental detection of spin textures}.---The topological properties
of 2D Rashba and 3D Weyl SOC can be characterized by their spin textures in
the momentum space, which are shown in Figs.~\ref{fig:detect}(a) and \ref%
{fig:detect}(b). These spin textures are determined by the effective field $%
\mathbf{h}_{\mathbf{k}}$ in the Hamiltonian Eq.~(\ref{Ham2}) and can be
parameterized by angles $\theta _{\mathbf{k}}=\arctan \left( \sqrt{h_{x%
\mathbf{k}}^{2}+h_{y\mathbf{k}}^{2}}/h_{z\mathbf{k}}\right) $ and $\varphi _{%
\mathbf{k}}=\arctan \left( h_{y\mathbf{k}}/h_{x\mathbf{k}}\right) $ at each
momentum $\mathbf{k}$. Here we propose a spectroscopic procedure \cite{supp}
to determine the parameter $\left( \theta _{\mathbf{k}},\varphi _{\mathbf{k}%
}\right) $ by measuring the time dynamics of spin polarization $\langle 
\mathbf{S}(\mathbf{k})\rangle $, which are obtained through a combination of
momentum-resolved Rabi spectroscopy \cite{clocktheory} and time-of-flight
imaging. Three spectroscopic sequences, as shown in Fig.~\ref{fig:detect}%
(c), are used to eliminate various side effects and obtain accurate results.
In all sequences, atoms are initially prepared in state $\left\vert
g\right\rangle $ at half filling without SOC.

I): First a $\pi $-pulse along $\sigma_{x}$ is applied using the clock
laser, which excites atoms from $|g\rangle $ to $|e\rangle $ 
%|\uparrow ,\mathbf{k}\rangle $ to $|\downarrow ,\mathbf{k+Q}\rangle$
with a momentum transfer $\mathbf{Q}=(k_{\mathrm{c}},k_{\mathrm{c}},k_{%
\mathrm{R}}^{z})$ that is the same as that of the SOC. Therefore the $\pi $%
-pulse couples two pseudospin states with the same quasi-momentum $|g,%
\mathbf{k}\rangle \leftrightarrow |e,\mathbf{k}\rangle $. Such a $\pi $%
-pulse can be implemented using one of the four SOC beams, with the other
three turned off by electro-optic amplitude modulators (EOAM)~\cite%
{yariv2007photonics}, as shown in Figs.~\ref{fig:experimental}(b) and \ref%
{fig:experimental}(c). In the quasi-momentum space, the tight-binding
dispersion of state $|e\rangle $ is inverted with respect to state $%
|g\rangle $, therefore the energy splitting at different $\mathbf{k}$ is
different. With suitably chosen clock laser frequency, we can selectively
excite atoms at certain $\mathbf{k}^{\ast }$ using a weak pulse (compared to
the tight-binding band width), such that only atoms near $\mathbf{k}^{\ast }$
are excited and atoms away from $\mathbf{k}^{\ast }$ are off-resonance and
remain in state $|g\rangle $. Notice that the resonance momentum $\mathbf{k}%
^{\ast }$ is not single valued, which form a circle in 2D and a surface
(spherical for type-I or ellipsoid-like for type-II Weyl points) in 3D. We
can select a different $\mathbf{k}^{\ast }$ by slightly changing the
frequency of the $\pi $-pulse, thus cover the whole momentum space.

After the $\pi $-pulse, we turn on the 2D (3D) SOC and the system evolves
under the Hamiltonian Eq.~(\ref{Ham2}) for an interval $\tau $. Then the SOC
and lattice potentials are turned off and the time-of-flight images are
taken to determine the spin polarization $\langle \mathbf{S}(\mathbf{k}%
)\rangle $ at each $\mathbf{k}$ on the $\mathbf{k}^{\ast }$ ring or surface
as a function of $\tau $. Suitable spin rotations using pulses along
different spin axis with clock lasers may be needed before the
time-of-flight to measure different components of $\langle \mathbf{S}(%
\mathbf{k})\rangle $.

II): Sequence II is the same as I except that the $\pi $-pulse is along $%
-\sigma _{x}$, which is used to eliminate effects caused by atoms near
resonance momenta $\mathbf{k}^{\ast }$ that may be partially pumped to $%
|e\rangle $ (with amplitude $f_{\mathbf{k}}$). %~\cite{supp}.
Because of the partial excitation amplitude $f_{\mathbf{k}}$, the spin
polarizations $\langle \mathbf{S}(\mathbf{k})\rangle _{\mathrm{I,II}}$ 
%\sim \bar{S}_i(f_k,\tau,\theta_k,\varphi_k)$
obtained from sequences I or II become complicated functionals of $f_{%
\mathbf{k}},\theta _{\mathbf{k}},\varphi _{\mathbf{k}},\tau $. However, $f_{%
\mathbf{k}}$ appears as a simple overall factor in their average~\cite{supp} 
\begin{equation}
\frac{\langle \mathbf{S(k)}\rangle _{\mathrm{I}}+\langle \mathbf{S(k)}%
\rangle _{\mathrm{II}}}{2} = (\frac{1}{2}-|f_{\mathbf{k}}|^{2})\mathbf{T}%
(\theta _{\mathbf{k}},\varphi _{\mathbf{k}},\tau ),  \label{eq:detect}
\end{equation}%
with $\mathbf{T}(\theta _{\mathbf{k}},\varphi _{\mathbf{k}},\tau )$ a simple
dynamical function which can be used to determine $\theta _{\mathbf{k}%
},\varphi _{\mathbf{k}}$.

III): Sequence III is the same as I without the $\pi $-pulse, which is used
to filter the dynamics of the excited atoms from the remaining $|g\rangle $
atoms by canceling the $1/2$ in Eq.~(\ref{eq:detect}). As a result, signals
for atoms with momenta far away from $\mathbf{k}^{\ast }$ are eliminated
because $f_{\mathbf{k}}$ is nonzero only in a narrow interval around $%
\mathbf{k}^{\ast }$ ring or surface. This process isolates the dynamics of
atoms near $\mathbf{k}^{\ast }$, and we can then replace $\theta _{\mathbf{k}%
},\varphi _{\mathbf{k}}$ by $\theta _{\mathbf{k^{\ast }}},\varphi _{\mathbf{%
k^{\ast }}}$ in Eq.~(\ref{eq:detect}), from which we can extract their
values~\cite{supp}.

\emph{Conclusions}.---In summary, we proposed a scheme for realizing and
detecting 2D Rashba and 3D Weyl types of SOC for alkaline-earth(-like) atoms
in optical lattice clocks without involving Raman process, therefore the
heating of atoms due to lasers is strongly suppressed. In combination with 
\textit{s}-wave scattering interaction between atoms, our scheme provides a
powerful platform for realizing stable topological superfluids and observing
associated topological excitations, such as Majorana fermions, which may
have potential applications in fault-tolerant topological quantum
computation.

\begin{acknowledgments}
\textbf{Acknowledgements}: X. Z., G. C., and S. J. are supported by National
Key R\&D Program of China under Grants No.~2017YFA0304203; the NSFC under
Grants No.~11434007 and No.~11674200; the PCSIRT under Grant No.~IRT13076;
SFSSSP; OYTPSP; and 1331KYC. X. L. and C. Z. are supported by AFOSR
(FA9550-16-1-0387), NSF (PHY-1505496), and ARO (W911NF-17-1-0128).
\end{acknowledgments}

%\clearpage
\onecolumngrid
\appendix

\section{Supplementary Materials}

In this supplemental materials we provide the derivation of the
tight-binding Hamiltonian, the phase diagram for $t\leq t_{z}$, 
%flat Fermi arc which connect the Weyl points
and some details of the detection scheme for spin textures.

\subsection{\textbf{A. Tight-binding Hamiltonian}}

After the single-band approximation and the unitary transformation $%
U=e^{-ik_{\mathrm{R}}^{z}z/2}\left\vert g\rangle \!\langle g\right\vert
+e^{ik_{\mathrm{R}}^{z}z/2}\left\vert e\rangle \!\langle e\right\vert $, we
obtain the tight-binding Hamiltonian from Eq.~(1) in the main text, which
can be written as 
\begin{eqnarray}
\mathcal{H}_{\mathrm{TI}} &=&-t\!\!\sum_{\left\langle \vec{j},\vec{j^{\prime
}}\right\rangle }\hat{c}_{\vec{j}s}^{\dag }\hat{c}_{\vec{j^{\prime }}%
s}+m_{z}\sum_{\vec{j}}\left( \hat{n}_{\vec{j}g}-\hat{n}_{\vec{j}e}\right) 
\notag \\
&&+\sum_{\left\langle \vec{j},\vec{j^{\prime }}\right\rangle }\left[ t_{%
\mathrm{so}}^{\vec{j},\vec{j^{\prime }}}\hat{c}_{\vec{j}g}^{\dag }\hat{c}_{%
\vec{j^{\prime }}e}+\mathrm{H.c.}\right]  \notag \\
&&-t_{z}\!\sum_{j_{z}}\left( e^{i\phi \xi _{s}/2}\hat{c}_{j_{z}s}^{\dag }%
\hat{c}_{j_{z}+1s}+\mathrm{H.c.}\right) ,  \label{HTI}
\end{eqnarray}%
where $\left\langle \ \right\rangle $ denotes the sum over nearest-neighbor
sites with $x$-$y$ plane lattice-site index $\vec{j}=(j_{x},j_{y})$,
particle number operators $\hat{n}_{\vec{j}s}=\hat{c}_{\vec{j}s}^{\dag }\hat{%
c}_{\vec{j}s}$, and effective Zeeman field $m_{z}=\hbar \delta $. $t$ is the
spin-preserved hopping amplitude along the $x$ ($y$)-direction, and 
%\begin{equation}
$t_{\mathrm{so}}^{\vec{j},\vec{j^{\prime }}}=\int
d^{2}rW_{g}(r-r_{j})MW_{e}(r-r_{j^{\prime }})$ %\end{equation}%
is the spin-flipped hopping parameter along the $x$ ($y$)-direction with $%
W_{g,e}(r-r_{j})$ the Wannier function at site $j$. Notice that the spatial
period of $M$ (in both $x$ and $y$-directions) is twice of the period of the
optical lattice. As a result, we have $t_{\mathrm{so}}^{\vec{j},\vec{j}\pm 
\vec{e}_{\mu }}=\pm (-1)^{j_{x}+j_{y}}t_{\mathrm{so}}^{\mu }$, with $\mu
=x,y $. For the aforementioned clock transition $M$, we have $t_{\mathrm{so}%
}^{x}=-it_{\mathrm{so}}^{y}=t_{\mathrm{so}}$. $t$ and $t_{\mathrm{so}}$ can
be tuned by changing the amplitude of clock laser's Rabi coupling $\Omega
_{0}$. The last term in Eq.~(\ref{HTI}) corresponds to spin conserved
hopping along the \textit{z}-direction, where $t_{z}$ is the hopping
amplitude, $\phi =\pi k_{\mathrm{R}}^{z}/k_{\mathrm{L}}^{z}$ is the hopping
phase (it can be tuned by the angle between the plane-wave laser pair that
forms the z-direction lattice), and $\xi _{g,e}=\pm 1$. Due to the spatial
dependence of the SOC, the Hamiltonian of $x$-$y$ plane have a period $2a$,
which can be restore to $a$ by applying the following unitary transformation 
$\hat{c}_{\vec{j}e}\longrightarrow e^{i\pi \left( x_{j}+y_{j}\right) /a}\hat{%
c}_{\vec{j}e}$. The Fourier transformation of the Eq.~(\ref{HTI}) to
momentum space yield the momentum effective Hamiltonian Eq.~(2) in the main
text.

\subsection{\textbf{B. Phase diagram for $t\leq t_{z}$}}

\begin{figure}[t]
\includegraphics[width = 2.0in]{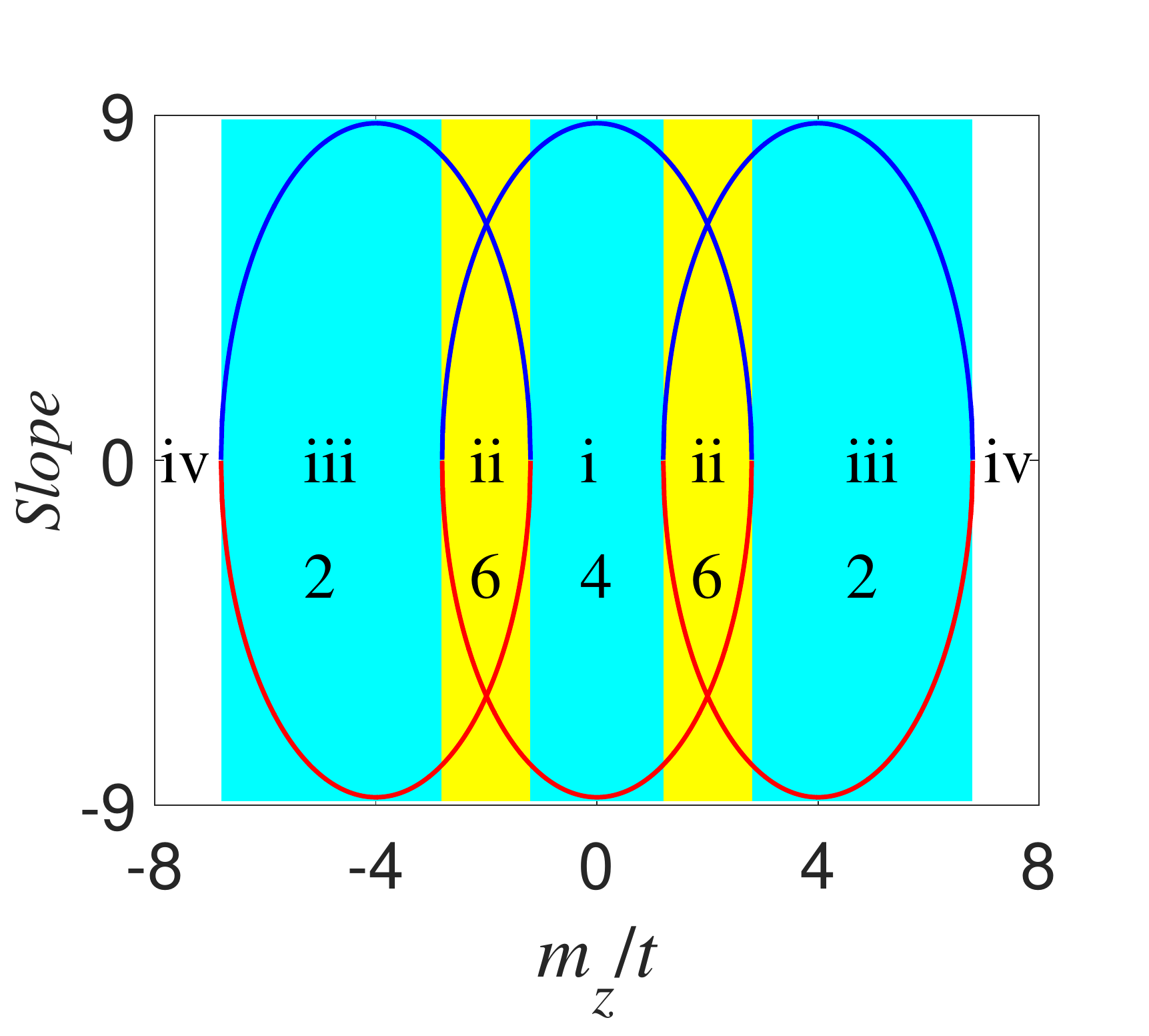} \hskip -0.0cm 
\centering
\caption{The phase diagram with Weyl type of SOC. $\protect\phi =\protect\pi 
$, $t_{\mathrm{so}}/t=0.6$ and $t_{z}/t=1.4$. $(2,6,4,6,2)$ indicate the
number of Weyl points. Blue and red lines represent two slopes $\pm
2t_{z}\cos \!\left( \protect\gamma \right) $ of the energy dispersions along
the $k_{z}$ direction near a Weyl point.}
\label{fig:phase_diagram_mz_tsz}
\end{figure}

In the main text, we have discussed the topological phases with 3D SOC for $%
t>t_{z}$. For $t\leq t_{z}$, the physics is similar except that more Weyl
points may be found in certain parameter region. 
%The phase diagram for $t = t_{z}$ with $\phi =\pi $ is same as that of $t > t_{z}$ except that
%there is no $ii$ phase.
For simplicity, here we only consider type-I Weyl points with $\phi =\pi $ (%
\textit{i.e.}, $h_{0\mathbf{k}}=0$). As we discussed in the main text, 
%From the energy dispersion $E_{\mathbf{k}}=\pm |\mathbf{h}_{%
%\mathbf{k}}|$, we see the spectrum is gapless only when $|\mathbf{h}_{%
%\mathbf{k}}|=0$, with $h_{x\mathbf{k}}=h_{y\mathbf{k}}=0$
the Weyl points can only occur at $k_{x}a,k_{y}a=0$ or $\pi $. For $\left(
k_{x}a,k_{y}a\right) =\left( 0,\pi \right) $ or $\left( \pi ,0\right) $, $%
\mathbf{h}_{\mathbf{k}}$ becomes zero at $k_{z}a_{z}=\gamma $ or $-\pi
\!-\!\gamma $ with $\gamma =\arcsin \!\left[ \frac{-m_{z}}{2t_{z}}\right] $,
leading to 4 Weyl points in the region $\left\vert m_{z}\right\vert \leq
2t_{z}$. Similarly, we have 2 Weyl points in the region $4t-2t_{z}\leq
m_{z}< 4t+2t_{z}$ ($-4t-2t_{z}< m_{z}\leq -4t+2t_{z}$) for $\left(
k_{x}a,k_{y}a\right) =\left( 0,0\right) $ [$\left( k_{x}a,k_{y}a\right)
=\left( \pi ,\pi \right) $]. For $t_{z}<t$, all these regions are separated
without any overlap, therefore we have two fully gapped regions without any
Weyl point in $2t_{z}<|m_{z}|<4t-2t_{z}$. While for $2t\geq t_{z}>t$, the
region $\left\vert m_{z}\right\vert \leq 2t_{z}$ overlaps with regions $%
4t-2t_{z}\leq m_{z}< 4t+2t_{z}$ and $-4t-2t_{z}< m_{z}\leq -4t+2t_{z}$,
leading to 6 Weyl points in the regions $4t-2t_{z}\leq m_{z}\leq 2t_{z}$ and 
$-2t_{z}\leq m_{z}\leq 2t_{z}-4t$, as shown in Fig.~\ref%
{fig:phase_diagram_mz_tsz}. Moreover, for $t_{z}>2t$, all these three
regions overlaps, leading to a new region $4t-2t_{z}\leq m_{z}\leq 2t_{z}-4t$
with 8 Weyl points.

\subsection{\textbf{C. Experimental detection of spin textures using three
spectroscopic sequences}}

Before introducing the details of our detection scheme, we first show how
the spin polarization evolves under the SOC Hamiltonian. Consider an atom in
an arbitrary initial state $|\psi _{0}(\mathbf{k})\rangle =\alpha |g_{%
\mathbf{k}}\rangle +\beta |e_{\mathbf{k}}\rangle $ with momentum $\mathbf{k}$%
. Under the SOC Hamiltonian, the state becomes 
\begin{eqnarray}
|\psi _{\tau }(\mathbf{k})\rangle &=&\left[ \left( \alpha \frac{\cos \theta
_{\mathbf{k}}+1}{2}+\beta \frac{\sin \theta _{\mathbf{k}}}{2}e^{-i\varphi _{%
\mathbf{k}}}\right) e^{-iE_{\mathbf{k},-}\tau }+\left( \alpha \frac{1-\cos
\theta _{\mathbf{k}}}{2}-\beta \frac{\sin \theta _{\mathbf{k}}}{2}%
e^{-i\varphi _{\mathbf{k}}}\right) e^{-iE_{\mathbf{k},+}\tau }\right] |g_{%
\mathbf{k}}\rangle  \notag \\
&&+\left[ \left( \alpha \frac{\sin \theta _{\mathbf{k}}}{2}e^{i\varphi _{%
\mathbf{k}}}+\beta \frac{1-\cos \theta _{\mathbf{k}}}{2}\right) e^{-iE_{%
\mathbf{k},-}\tau }-\left( \alpha \frac{\sin \theta _{\mathbf{k}}}{2}%
e^{i\varphi _{\mathbf{k}}}-\beta \frac{\cos \theta _{\mathbf{k}}+1}{2}%
\right) e^{-iE_{\mathbf{k},+}\tau }\right] |e_{\mathbf{k}}\rangle ,
\end{eqnarray}%
after time $\tau $ with $E_{\mathbf{k},\pm }=h_{0\mathbf{k}}\pm |\mathbf{h}_{%
\mathbf{k}}|$.

Due to the Rabi oscillation, the time dynamics of all components' spin
polarizations can be characterized as 
\begin{eqnarray}
\langle S_{z}\rangle &=&P_{z}\left( \alpha ,\beta ,\tau ,\mathbf{k}\right) 
\notag \\
&=&\frac{1}{2}\left( \left\vert \alpha \right\vert ^{2}-\left\vert \beta
\right\vert ^{2}\right) \left[ \cos ^{2}\theta _{\mathbf{k}}+\sin ^{2}\theta
_{\mathbf{k}}\cos \left( 2\left\vert \mathbf{h}_{\mathbf{k}}\right\vert \tau
\right) \right]  \notag \\
&&+\frac{1}{2}\left( \alpha \beta ^{\ast }+\alpha ^{\ast }\beta \right) \cos
\theta _{\mathbf{k}}\sin \theta _{\mathbf{k}}\cos \varphi _{\mathbf{k}}\left[
1-\cos \left( 2\left\vert \mathbf{h}_{\mathbf{k}}\right\vert \tau \right) %
\right]  \notag \\
&&+\left( -i\right) \frac{1}{2}\left( \alpha \beta ^{\ast }-\alpha ^{\ast
}\beta \right) \sin \theta _{\mathbf{k}}\cos \varphi _{\mathbf{k}}\sin
\left( 2\left\vert \mathbf{h}_{\mathbf{k}}\right\vert \tau \right) ,
\end{eqnarray}%
\begin{eqnarray}
\langle S_{y}\rangle &=&P_{y}\left( \alpha ,\beta ,\tau ,\mathbf{k}\right) 
\notag \\
&=&\frac{1}{4}\left( -i\right) \left( \alpha ^{\ast }\beta -\alpha \beta
^{\ast }\right) \left[ \sin ^{2}\theta _{\mathbf{k}}+\cos ^{2}\theta _{%
\mathbf{k}}\cos \left( 2\left\vert \mathbf{h}_{\mathbf{k}}\right\vert \tau
\right) \right]  \notag \\
&&+\frac{1}{2}\left( \left\vert \alpha \right\vert ^{2}-\left\vert \beta
\right\vert ^{2}\right) \cos \theta _{\mathbf{k}}\sin \theta _{\mathbf{k}%
}\sin \varphi _{\mathbf{k}}\left[ 1-\cos \left( 2\left\vert \mathbf{h}_{%
\mathbf{k}}\right\vert \tau \right) \right]  \notag \\
&&+\frac{1}{4}\left( -i\right) \left( \alpha \beta ^{\ast }-\alpha ^{\ast
}\beta \right) \sin ^{2}\theta _{\mathbf{k}}\cos 2\varphi _{\mathbf{k}}\left[
1-\cos \left[ 2\left\vert \mathbf{h}_{\mathbf{k}}\right\vert \tau \right] %
\right]  \notag \\
&&+\frac{1}{4}\left( \alpha \beta ^{\ast }+\alpha ^{\ast }\beta \right) \sin
^{2}\theta _{\mathbf{k}}\sin 2\varphi _{\mathbf{k}}\left[ 1-\cos \left(
2\left\vert \mathbf{h}_{\mathbf{k}}\right\vert \tau \right) \right]  \notag
\\
&&+\frac{1}{4}\left( -i\right) \alpha ^{\ast }\beta \cos \left( 2\left\vert 
\mathbf{h}_{\mathbf{k}}\right\vert \tau \right)  \notag \\
&&+\frac{1}{4}\left[ 2\left( \left\vert \alpha \right\vert ^{2}-\left\vert
\beta \right\vert ^{2}\right) \sin \theta _{\mathbf{k}}\cos \varphi _{%
\mathbf{k}}-2\left( \alpha ^{\ast }\beta +\alpha \beta ^{\ast }\right) \cos
\theta _{\mathbf{k}}-\alpha \beta ^{\ast }\right] \sin \left( 2\left\vert 
\mathbf{h}_{\mathbf{k}}\right\vert \tau \right) ,
\end{eqnarray}%
and 
\begin{eqnarray}
\langle S_{x}\rangle &=&P_{x}\left( \alpha ,\beta ,\tau ,\mathbf{k}\right) 
\notag \\
&=&\frac{1}{4}\left( \alpha ^{\ast }\beta +\alpha \beta ^{\ast }\right) %
\left[ \sin ^{2}\theta _{\mathbf{k}}+\cos ^{2}\theta _{\mathbf{k}}\cos
\left( 2\left\vert \mathbf{h}_{\mathbf{k}}\right\vert \tau \right) \right] 
\notag \\
&&+\frac{1}{2}\left( \left\vert \alpha \right\vert ^{2}-\left\vert \beta
\right\vert ^{2}\right) \cos \theta _{\mathbf{k}}\sin \theta _{\mathbf{k}%
}\cos \varphi _{\mathbf{k}}\left[ 1-\cos \left( 2\left\vert \mathbf{h}_{%
\mathbf{k}}\right\vert \tau \right) \right]  \notag \\
&&+\frac{1}{4}\left( \alpha \beta ^{\ast }+\alpha ^{\ast }\beta \right) \sin
^{2}\theta _{\mathbf{k}}\cos 2\varphi _{\mathbf{k}}\left[ 1-\cos \left(
2\left\vert \mathbf{h}_{\mathbf{k}}\right\vert \tau \right) \right]  \notag
\\
&&+\frac{1}{4}i\left( \alpha \beta ^{\ast }-\alpha ^{\ast }\beta \right)
\sin ^{2}\theta _{\mathbf{k}}\sin 2\varphi _{\mathbf{k}}\left[ 1-\cos \left(
2\left\vert \mathbf{h}_{\mathbf{k}}\right\vert \tau \right) \right]  \notag
\\
&&+\frac{1}{4}\alpha ^{\ast }\beta \cos \left( 2\left\vert \mathbf{h}_{%
\mathbf{k}}\right\vert \tau \right)  \notag \\
&&+\frac{1}{4}\left[ 2\left( \left\vert \alpha \right\vert ^{2}-\left\vert
\beta \right\vert ^{2}\right) \sin \theta _{\mathbf{k}}i\sin \varphi _{%
\mathbf{k}}-2\left( \alpha ^{\ast }\beta -\alpha \beta ^{\ast }\right) \cos
\theta _{\mathbf{k}}+\alpha \beta ^{\ast }\right] i\sin \left( 2\left\vert 
\mathbf{h}_{\mathbf{k}}\right\vert \tau \right) .
\end{eqnarray}%
These equations show the relations between the angles $\left( \varphi _{%
\mathbf{k}},\theta _{\mathbf{k}}\right) $ and the dynamics of spin
polarization. In the following, we show that these relations can used to
determine $\left( \varphi _{\mathbf{k}},\theta _{\mathbf{k}}\right) $
through three spectroscopic sequences (I, II and III), as shown in Fig.~4(c)
in the main text. In all sequences, atoms are initially prepared in the
state $|g\rangle $ at half filling without SOC.

%In experimental, three spectroscopic sequences (I, II and III)
%are used to detect the dynamic of spin polarization. Every sequence follows
%the steps which contains preparing the atoms in $|g\rangle $, applying a $%
%\pi $-pulse, turning on the 2D(3D) SOC for an interval $\tau $,
%time-of-flight imaging, and finally measure $\mathbf{S}$, as shown in
%Fig.~4(c) in the main text.

In sequence I, a $\pi $-pulse along $\sigma _{x}$ is applied using the clock
laser to induce the $|g\rangle \rightarrow |e\rangle $ transition at the
same quasi-momentum. Such a $\pi $-pulse can be implemented using one of the
four SOC beams, with the other three turned off by electro-optic amplitude
modulators. We can selectively excite atoms at certain $\mathbf{k}^{\ast }$
using a weak pulse (compared to the tight-binding band width), such that
only atoms near $\mathbf{k}^{\ast }$ are excited to $|e\rangle $ 
%(with amplitude $f_\mathbf{k}$)
and atoms away from $\mathbf{k}^{\ast }$ are off-resonance and remain in
state $|g\rangle $. Then we turn on the 2D (3D) SOC for an interval $\tau $
and let the system evolve under the Hamiltonian Eq.~(2) in the main text.
Even for a weak pulse, atoms near resonance momenta $\mathbf{k}^{\ast }$ may
still be partially pumped to $|e\rangle $ (with amplitude $f_{\mathbf{k}}$). 
%~\cite{supp}.
Because of the partial excitation amplitude $f_{\mathbf{k}}$, the spin
polarization becomes a complex function of $\sqrt{1-|f_{\mathbf{k}}|^{2}}%
,-if_{\mathbf{k}},\tau ,\mathbf{k}$, which is $\langle \mathbf{S(\mathbf{k})}%
\rangle _{\text{I}}=\mathbf{P}\left( \sqrt{1-|f_{\mathbf{k}}|^{2}},-if_{%
\mathbf{k}},\tau ,\mathbf{k}\right) $.

To eliminate the effects of partial excitation, we introduce sequence II
which is the same as I except that the $\pi $-pulse is along $-\sigma _{x}$,
leading to the spin polarization $\langle \mathbf{S(\mathbf{k})}\rangle _{%
\text{II}}=\mathbf{P}\left( \sqrt{1-|f_{\mathbf{k}}|^{2}},if_{\mathbf{k}%
},\tau ,\mathbf{k}\right) $. The average between these two sequences gives a
simple form of the spin polarization as 
\begin{equation}
\frac{1}{2}\left[ \langle \mathbf{S(\mathbf{k})}\rangle _{\text{I}}+\langle 
\mathbf{S(\mathbf{k})}\rangle _{\text{II}}\right] =\frac{1}{2}\left( \mathbf{%
P}(\sqrt{1-|f_{\mathbf{k}}|^{2}},-if_{\mathbf{k}},\tau ,\mathbf{k})+\mathbf{P%
}(\sqrt{1-|f_{\mathbf{k}}|^{2}},if_{\mathbf{k}},\tau ,\mathbf{k})\right) =(%
\frac{1}{2}-|f_{\mathbf{k}}|^{2})\mathbf{T}(\theta _{\mathbf{k}},\varphi _{%
\mathbf{k}},\tau ).  \label{eq:avg}
\end{equation}%
Here $\mathbf{T}$ is given as 
\begin{eqnarray*}
T_{z}\left( \theta _{\mathbf{k}},\varphi _{\mathbf{k}},\tau \right) &=&\left[
\cos ^{2}\theta _{\mathbf{k}}+\sin ^{2}\theta _{\mathbf{k}}\cos \left( 2|%
\mathbf{h}_{\mathbf{k}}|\tau \right) \right] , \\
T_{y}\left( \theta _{\mathbf{k}},\varphi _{\mathbf{k}},\tau \right)
&=&\left\{ \cos \theta _{\mathbf{k}}\sin \theta _{\mathbf{k}}\sin \varphi _{%
\mathbf{k}}-\Theta \sin \left[ 2|\mathbf{h}_{\mathbf{k}}|\tau +\arctan
\left( -\cos \theta _{\mathbf{k}}\tan \varphi _{\mathbf{k}}\right) \right]
\right\} , \\
T_{x}\left( \theta _{\mathbf{k}},\varphi _{\mathbf{k}},\tau \right)
&=&\left\{ \cos \theta _{\mathbf{k}}\sin \theta _{\mathbf{k}}\cos \varphi _{%
\mathbf{k}}-\Gamma \sin \left[ 2|\mathbf{h}_{\mathbf{k}}|\tau +\arctan
\left( \frac{\cos \theta _{\mathbf{k}}}{\tan \varphi _{\mathbf{k}}}\right) %
\right] \right\} ,
\end{eqnarray*}%
with $\Theta =\sqrt{\left( \cos \theta _{\left\vert \mathbf{k}\right\vert
}\sin \theta _{\left\vert \mathbf{k}\right\vert }\sin \varphi _{\left\vert 
\mathbf{k}\right\vert }\right) ^{2}+\left( \sin \theta _{\left\vert \mathbf{k%
}\right\vert }\cos \varphi _{\left\vert \mathbf{k}\right\vert }\right) ^{2}}$%
, and $\Gamma =\sqrt{\left( \cos \theta _{\left\vert \mathbf{k}\right\vert
}\sin \theta _{\left\vert \mathbf{k}\right\vert }\cos \varphi _{\left\vert 
\mathbf{k}\right\vert }\right) ^{2}+\left( \sin \theta _{\left\vert \mathbf{k%
}\right\vert }\sin \varphi _{\left\vert \mathbf{k}\right\vert }\right) ^{2}}$%
.

Finally, we use sequence III (which is the same as I without the $\pi $%
-pulse) to filter the dynamics of the excited atoms from the remaining $%
|g\rangle $ atoms by canceling the $1/2$ in Eq.~(\ref{eq:avg}), yielding 
\begin{equation}
\frac{1}{2}\left[ \langle \mathbf{S(\mathbf{k})}\rangle _{\text{I}}+\langle 
\mathbf{S(\mathbf{k})}\rangle _{\text{II}}\right] -\langle \mathbf{S(\mathbf{%
k})}\rangle _{\text{III}}=|f_{\mathbf{k}}|^{2}\mathbf{T}(\theta _{\mathbf{k}%
},\varphi _{\mathbf{k}},\tau ).  \label{eq:123}
\end{equation}

\begin{figure}[t]
\includegraphics[width = 3.6in]{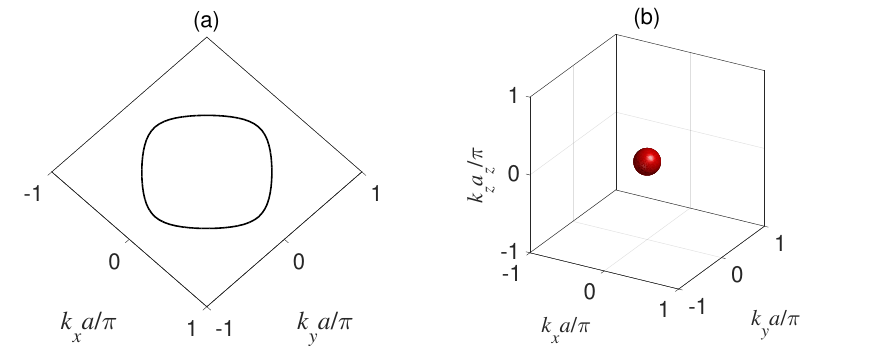} \hskip -0.0cm \centering
\caption{The resonance momenta $\mathbf{k}^{\ast }$ induced by a narrow $%
\protect\pi $-pulse is a circle in 2D (a) and a spherical surface in 3D (b)
which surrounds a Weyl point.}
\label{fig:k_2D_3D}
\end{figure}

\begin{figure}[t]
\includegraphics[width = 6.0in]{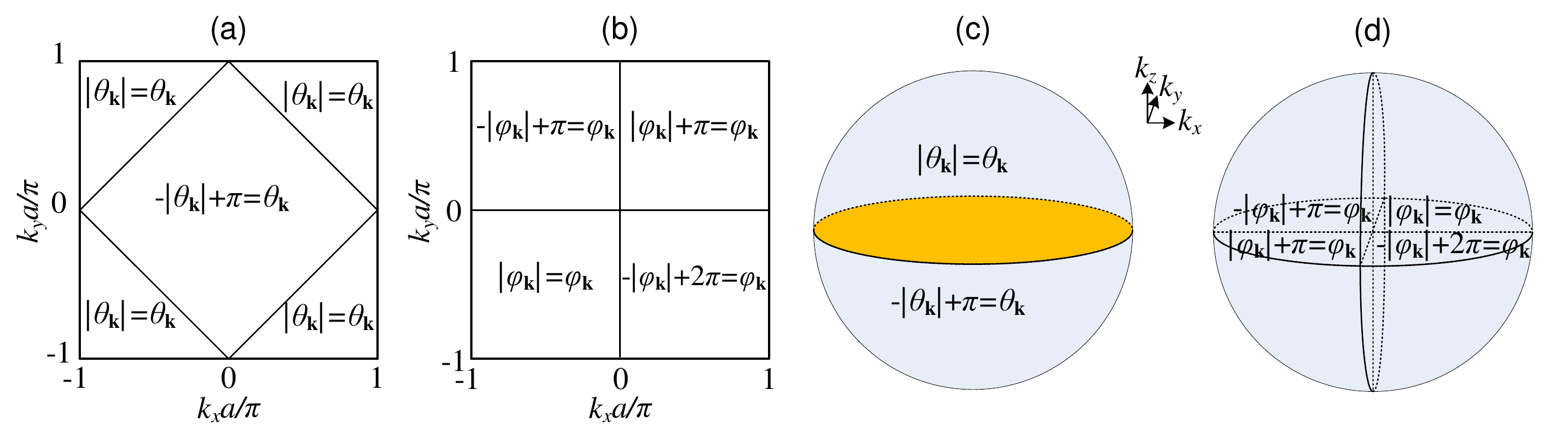} \hskip -0.0cm \centering
\caption{(a) and (b) The relations between $|\protect\theta _{\mathbf{k}}|$
and $\protect\theta _{\mathbf{k}}$, $|\protect\varphi _{\mathbf{k}}|$ and $%
\protect\varphi _{\mathbf{k}}$ in 2D. (c) and (d) The relations between $|%
\protect\theta _{\mathbf{k}}|$ and $\protect\theta _{\mathbf{k}}$, $|\protect%
\varphi _{\mathbf{k}}|$ and $\protect\varphi _{\mathbf{k}}$ near $+1$ Weyl
points in 3D.}
\label{fig:theta_phi_23D}
\end{figure}

Without the time-of-flight imaging, the experimentally measured
spin-polarization should be summed over the momentum space. We notice that $%
f_{\mathbf{k}}$ in Eq.~(\ref{eq:123}) is only nonzero around the $\mathbf{k}%
^{\ast }$ ring or surface (see Figs.~\ref{fig:k_2D_3D}). As a result,
time-of-flight images are taken to determine the spin polarization $\langle 
\mathbf{S}\rangle $ at each $\mathbf{k}$ only on the $\mathbf{k}^{\ast }$
ring or surface, rather than the whole momentum space. That is to say, we
use the time-of-flight to resolve the azimuthal direction of momentum, and
the final observable for a certain point $\mathbf{k}^{\ast }$ is 
\begin{equation}
\mathbf{O}(\mathbf{k}^{\ast },\tau )=\int |f_{\mathbf{k}}|^{2}\mathbf{T}%
(\theta _{\mathbf{k}},\varphi _{\mathbf{k}},\tau )d|\mathbf{k}|\simeq c%
\mathbf{T}(\theta _{\mathbf{k}^{\ast }},\varphi _{\mathbf{k}^{\ast }},\tau ).
\label{OA}
\end{equation}%
Here $c=\int |f_{\mathbf{k}}|^{2}d|\mathbf{k}|$ is a constant (which is
integrated in the radial direction) and we have taken into account that $f_{%
\mathbf{k}}$ is only nonzero around the $\mathbf{k}^{\ast }$. The angles $%
|\theta _{\mathbf{k}^{\ast }}|$ and $|\phi _{\mathbf{k}^{\ast }}|$ can be
inferred from the oscillations of $\mathbf{O}(\mathbf{k}^{\ast },\tau )$: 
\begin{eqnarray}
2\tan ^{2}\theta _{\mathbf{k}^{\ast }} &=&\frac{O_{z}(\mathbf{k}^{\ast
},\tau )_{\mbox{max}}-O_{z}(\mathbf{k}^{\ast },\tau )_{\mbox{min}}}{O_{z}(%
\mathbf{k}^{\ast },\tau )_{\mbox{mean}}},  \notag \\
2\sqrt{1+\frac{1}{\cos ^{2}\theta _{\mathbf{k}^{\ast }}\tan ^{2}\varphi _{%
\mathbf{k}^{\ast }}}} &=&\frac{O_{y}(\mathbf{k}^{\ast },\tau )_{\mbox{max}%
}-O_{y}(\mathbf{k}^{\ast },\tau )_{\mbox{min}}}{O_{y}(\mathbf{k}^{\ast
},\tau )_{\mbox{mean}}},  \notag \\
2\sqrt{1+\frac{\tan ^{2}\varphi _{\mathbf{k}^{\ast }}}{\sin ^{2}\theta _{%
\mathbf{k}^{\ast }}}} &=&\frac{O_{x}(\mathbf{k}^{\ast },\tau )_{\mbox{max}%
}-O_{x}(\mathbf{k}^{\ast },\tau )_{\mbox{min}}}{O_{x}(\mathbf{k}^{\ast
},\tau )_{\mbox{mean}}}.
\end{eqnarray}%
$\theta _{\mathbf{k}^{\ast }}$ and $\varphi _{\mathbf{k}^{\ast }}$ are
determined using the principle of continuity in the momentum space, as shown
in Fig.~\ref{fig:theta_phi_23D}. Changing the frequency of the $\pi $-pulse,
we can obtain $\theta _{\mathbf{k}}$ and $\phi _{\mathbf{k}}$ in the whole
momentum space.

\begin{thebibliography}{99}
\bibitem{Xiao} D. Xiao, M.-C. Chang, and Q. Niu, Berry phases effects on
electronic properties, Rev. Mod. Phys. \textbf{82}, 1959 (2010).

\bibitem{Hasan} M. Z. Hasan and C. L. Kane, Topological insulators, Rev.
Mod. Phys. \textbf{82}, 3045 (2010).

\bibitem{Qi} X.-L. Qi and S.-C. Zhang, Topological insulators and
superconductors, Rev. Mod. Phys. \textbf{83}, 1057 (2011).

\bibitem{Moore} J. E. Moore, The birth of topological insulators, Nature
(London) \textbf{464}, 194 (2010).

\bibitem{spielman} Y.-J. Lin, K. Jim\'{e}nez-Garc\'{\i}a, and I. B.
Spielman, Spin-orbit-coupled Bose-Einstein condensates, Nature (London) 
\textbf{471}, 83 (2011).

\bibitem{FuPRA} Z. Fu, P. Wang, S. Chai, L. Huang, and J. Zhang, Bose-
Einstein condensate in a light-induced vector gauge potential using the 1064
nm optical dipole trap lasers, Phys. Rev. A \textbf{84}, 043609 (2011).

\bibitem{Shuai-PRL} J.-Y. Zhang, S.-C. Ji, Z. Chen, L. Zhang, Z.-D. Du, B.
Yan, G.-S. Pan, B. Zhao, Y.-J. Deng, H. Zhai, S. Chen, and J.-W. Pan,
Collective dipole oscillations of a spin-orbit coupled Bose-Einstein
condensate, Phys. Rev. Lett. \textbf{109}, 115301 (2012).

\bibitem{Washington-PRA} C. Qu, C. Hamner, M. Gong, C. Zhang, and P. Engels,
Observation of Zitterbewegung in a spin-orbit coupled Bose-Einstein
condensates, Phys. Rev. A \textbf{88}, 021604(R) (2013)

\bibitem{Purdue} A. J. Olson, S.-J. Wang, R. J. Niffenegger, C. -H. Li, C.
H. Greene, and Y. P. Chen, Tunable Laudan-Zener transitions in a
spin-orbit-coupled Bose-Einstein condesate, Phys. Rev. A \textbf{90}, 013616
(2014).

\bibitem{Jing} P. Wang, Z. Yu, Z. Fu, J. Miao, L. Huang, S. Chai, H. Zhai,
and J. Zhang, Spin-orbit coupled degenerate Fermi gases, Phys. Rev. Lett. 
\textbf{109}, 095301 (2012).

\bibitem{MIT} L. W. Cheuk, A. T. Sommer, Z. Hadzibabic, T. Yefsah, W. S.
Bakr, and M. W. Zwierlein, Spin-injection spectroscopy of a spin-orbit
coupled Fermi gas, Phys. Rev. Lett. \textbf{109}, 095302 (2012).

\bibitem{Spielman-Fermi} R. A. Williams, M. C. Beeler, L. J. LeBlanc, and I.
B. Spielman, Raman-induced interactions in a single-component Fermi gas near
an \textit{s}-wave Feshbach resonance, Phys. Rev. Lett. \textbf{111}, 095301
(2013).

\bibitem{2DSOChuang} L. Huang, Z. Meng, P. Wang, P. Peng, S.-L. Zhang, L.
Chen, D. Li, Q. Zhou, and J. Zhang, Experimental realization of
two-dimensional synthetic spin-orbit coupling in ultracold Fermi gases, Nat.
Phys. \textbf{12}, 540 (2016).

\bibitem{2DSOCpan} Z. Wu, L. Zhang, W. Sun, X.-T. Xu, B.-Z. Wang, S.-C. Ji,
Y. Deng, S. Chen, X.-J. Liu, and J.-W. Pan, Realization of two-dimensional
spin-orbit coupling for Bose-Einstein condensates, Science \textbf{354}, 83
(2016).

\bibitem{2DSOCmeng} Z. Meng, L. Huang, P. Peng, D. Li, L. Chen, Y. Xu, C.
Zhang, P. Wang, and J. Zhang, Experimental observation of a topological band
gap opening in ultracold Fermi gases with two-dimensional spin-orbit
coupling, Phys. Rev. Lett. \textbf{117}, 235304 (2016).

\bibitem{Majorana2008} C. Nayak, S. H. Simon, A. Stern, M. Freedman, and S.
D. Sarma, Non-Abelian anyons and topological quantum computation, Rev. Mod.
Phys. \textbf{80}, 1083 (2008).

\bibitem{Majorana2009} F. Wilczek, Majorana returns, Nat. Phys. \textbf{5},
614 (2009).

\bibitem{Zhang2008} C. Zhang, S. Tewari, R. M. Lutchyn, and S. Das Sarma, $%
p_{x}+ip_{y}$ superfluid from \textit{s}-wave interactions of fermionic cold
atoms, Phys. Rev. Lett. \textbf{101}, 160401 (2008).

\bibitem{Majorana2012} M. Gong, G. Chen, S. Jia, and C. Zhang, Searching for
Majorana fermions in 2D spin-orbit coupled Fermi superfluids at finite
temperature, Phys. Rev Lett. \textbf{109}, 105302 (2012).

\bibitem{Weyl1} M. Gong, S. Tewari, and C. Zhang, BCS-BEC crossover and
topological phase transition in 3D spin-orbit coupled degenerate Fermi
gases, Phys. Rev. Lett. \textbf{107}, 195303 (2011).

\bibitem{Weyl2} Y. Xu, F. Zhang, and C. Zhang, Structured Weyl points in
spin-orbit coupled fermionic superfluids, Phys. Rev. Lett. \textbf{115},
265304 (2015).

\bibitem{Weyl3} Y. Xu and L.-M. Duan, Type-II Weyl points in
three-dimensional cold-atom optical lattices, Phys. Rev. A \textbf{94},
053619 (2016).

\bibitem{Weyl4} T. Dub\v{c}ek, C. J. Kennedy, L. Lu, W. Ketterle, M. Solja%
\v{c}i\'{c}, and H. Buljan, Weyl points in three-dimensional optical
lattices: synthetic magnetic monopoles in momentum space, Phys. Rev. Lett. 
\textbf{114}, 225301 (2015).

\bibitem{Weyl5} B.-Z. Wang, Y.-H. Lu, W. Sun, S. Chen, Y. Deng, and X.-J.
Liu, Dirac, Rashba and Weyl type spin-orbit couplings: toward experimental
realization in ultracold atoms, Phys. Rev. A \textbf{97}, 011605 (2018).

\bibitem{Monopole} G. E. Volovik, 
\newblock {\em The universe in a helium
droplet} (Clarendon Press, Oxford, 2003).

\bibitem{lanth1} X. Cui, B. Lian, T.-L. Ho, B. L. Lev, and H. Zhai,
Synthetic gauge field with highly magnetic lanthanide atoms, Phys. Rev. A 
\textbf{88}, 011601 (2013).

\bibitem{lanth2} N. Q. Burdick, Y. Tang, and B. L. Lev, Long-lived
spin-orbit-coupled degenerate dipolar Fermi gas, Phys. Rev. X \textbf{6},
031022 (2016).

\bibitem{clocktheory} M. L. Wall, A. P. Koller, S. Li, X. Zhang, N. R.
Cooper, J. Ye, and A. M. Rey, Synthetic spin-orbit coupling in an optical
lattice clock, Phys. Rev. Lett. \textbf{116}, 035301 (2016).

\bibitem{clock173Yb} L. F. Livi, G. Cappellini, M. Diem, L. Franchi, C.
Clivati, M. Frittelli, F. Levi, D. Calonico, J. Catani, M. Inguscio, and L.
Fallani, Synthetic dimensions and spin-orbit coupling with an optical clock
transition, Phys. Rev. Lett. \textbf{117}, 220401 (2016).

\bibitem{clock87Sr} S. Kolkowitz, S. L. Bromley, T. Bothwell, M. L. Wall, G.
E. Marti, A. P. Koller, X. Zhang, A. M. Rey, and J. Ye, Spin-orbit-coupled
fermions in an optical lattice clock, Nature (London) \textbf{542}, 66
(2017).

\bibitem{clocksoc} S. L. Bromley, S. Kolkowitz, T. Bothwell, D. Kedar, A.
Safavi-Naini, M.L. Wall, C. Salomon, A.M. Rey, and J. Ye, Dynamics of
interacting fermions under spin-orbit coupling in an optical lattice clock,
Nat. Phys. \textbf{14}, 399 (2018).

\bibitem{3Dclockbose} T. Akatsuka, M. Takamoto, and H. Katori,
Three-dimensional optical lattice clock with bosonic $^{88}$Sr atoms, Phys.
Rev. A \textbf{81}, 023402 (2010).

\bibitem{3Dclockfermi} S. L. Campbell, R. B. Hutson, G. E. Marti, A. Goban,
N. Darkwah Oppong, R. L. McNally, L. Sonderhouse, J. M. Robinson, W. Zhang,
B. J. Bloom, and J. Ye, A Fermi-degenerate three-dimensional optical lattice
clock, Science \textbf{358}, 90 (2017).

\bibitem{clockreview} A. D. Ludlow, M. M. Boyd, J. Ye, E. Peik, and P. O.
Schmidt, Optical atomic clocks, Rev. Mod. Phys. \textbf{87}, 637 (2015).

\bibitem{typeII} A. A. Soluyanov, D. Gresch, Z. Wang, Q. Wu, M. Troyer, X.
Dai, and B. A. Bernevig, Type-II Weyl semimetals, Nature (London) \textbf{527%
}, 495 (2015).

\bibitem{supp} See Supplemental Materials, for the derivation of the
tight-binding Hamiltonian, the phase diagram for $t\leq t_{z}$, and some
details of the detection scheme for spin textures.

\bibitem{Xiongjun2014} X.-J. Liu, K. T. Law, and T. K. Ng, Realization of 2D
spin-orbit interaction and exotic topological orders in cold atoms, Phys.
Rev. Lett. \textbf{112}, 086401 (2014).

\bibitem{Lifshitz} I. M. Lifshitz, Anomalies of electron characteristics of
a metal in the high pressure region, Sov. Phys. JETP \textbf{11}, 1130
(1960).

\bibitem{yariv2007photonics} A.~Yariv and P.~Yeh, 
\newblock {\em Photonics:
Optical Electronics in Modern Communications} (Oxford University Press,
Oxford, 2007).

%\bibitem{Yanreview} B. Yan and C. Felser, Topological Materials: Weyl Semimetals, Annu. Rev. Condens. Matter Phys. \textbf{8}, 337 (2017).

%\bibitem{photoemissionspectroscopy2008} J. T. Stewart, J. P. Gaebler, and D.
%S. Jin, Using photoemission spectroscopy to probe a strongly interacting
%Fermi gas, Nature (London) \textbf{454}, 744 (2008)
%
%\bibitem{detection1} I. Bloch, J. Dalibard, and W. Zwerger, Many-body
%physics with ultracold gases, Rev. Mod. Phys. \textbf{80}, 885 (2008).
%
%\bibitem{detection2} Ph. T. Ernst, S. Gtze, J. S. Krauser, K. Pyka,
%Dirk-Soren Lhmann, D. Pfannkuche, and K. Sengstock, Probing superfluids in
%optical lattices by momentum-resolved Bragg spectroscopy, Nat. Phys. \textbf{%
%6}, 56 (2010).
\end{thebibliography}
\end{document}